\documentclass[preprint]{elsarticle}

\usepackage{euscript,amssymb,amsmath}

\usepackage{graphicx}
\usepackage{epsfig}
\usepackage{color}

\newcommand{\be}[1]{\begin{equation}\label{#1}}
\newcommand{\ee}{\end{equation}}
\newcommand{\ba}[1]{\begin{eqnarray}\label{#1}}
\newcommand{\ea}{\end{eqnarray}}

\begin{document}

\begin{frontmatter}

\title{ {\bf Revolution of S-stars and oscillation\\ of solar and terrestrial observables:\\ nonrandom coincidence of periods} }

\author[a]{V.D.~Rusov}\ead{siiis@te.net.ua}

\author[a]{V.P.~Smolyar}\ead{vladimirsmolyar@ukr.net}

\author[b]{M.V.~Eingorn}\ead{maxim.eingorn@gmail.com}

\address[a]{Department of Theoretical and Experimental Nuclear Physics,\\ Odessa National Polytechnic University, Odessa, Ukraine\\}

\address[b]{CREST and NASA Research Centers,\\ North Carolina Central University, Durham, North Carolina, U.S.A.\\}

\begin{abstract}
A striking coincidence of revolution periods of S-stars orbiting a supermassive black hole at the Galactic Center of the Milky Way and oscillation periods of
such solar and terrestrial observables as the sunspot number, the geomagnetic field Y-component and the global temperature is established on basis of the
corresponding experimental data. Rejecting randomness of this discovered coincidence, we put forward a hypothesis that modulation of dark matter flows in the
Milky Way by the S-stars is responsible for such a frequency transfer from the Galactic Center to the Solar System.
\end{abstract}


\end{frontmatter}

\

\


Inspired by the surprising idea that the dark matter flux in the Solar System particularly and in the Milky Way generally can be modulated with the periods
equal to those of the solar cycles (including the 11-year one, being also typical for the cosmic muon flux measured at the Gran Sasso laboratory
\cite{Fernandez}), actually causing these cycles themselves, in this brief Letter we find a source for such a modulation in the vicinity of the Galactic Center
and establish a connection of this source (namely, S-stars) with some selected solar and terrestrial observables.


Let us address the well-known experimental data on the oscillations of the sunspot number, the geomagnetic field Y-component and the global temperature (Fig.
1, 2) and try to answer the following sudden and unconventional question: is there any observed relationship between their periods and revolution periods of
S-stars orbiting a supermassive black hole at the Galactic Center of the Milky Way?

\

In Fig. 1 we depict the power spectra of the sunspot number (1874-2010, the Royal Greenwich Observatory), the geomagnetic field Y-component (Greenwich and
Eskdalemuir observatories) and HadCRUT4 GST (1850-2012). NH and SH stand for Northern Hemisphere and Southern Hemisphere respectively, while MESA indicates the
maximum entropy spectral analysis. Pink stripes correspond to the major heliospheric harmonics.

\

\begin{figure}[htbp]
\begin{center}\includegraphics[width=4.85in,height=5.05in]{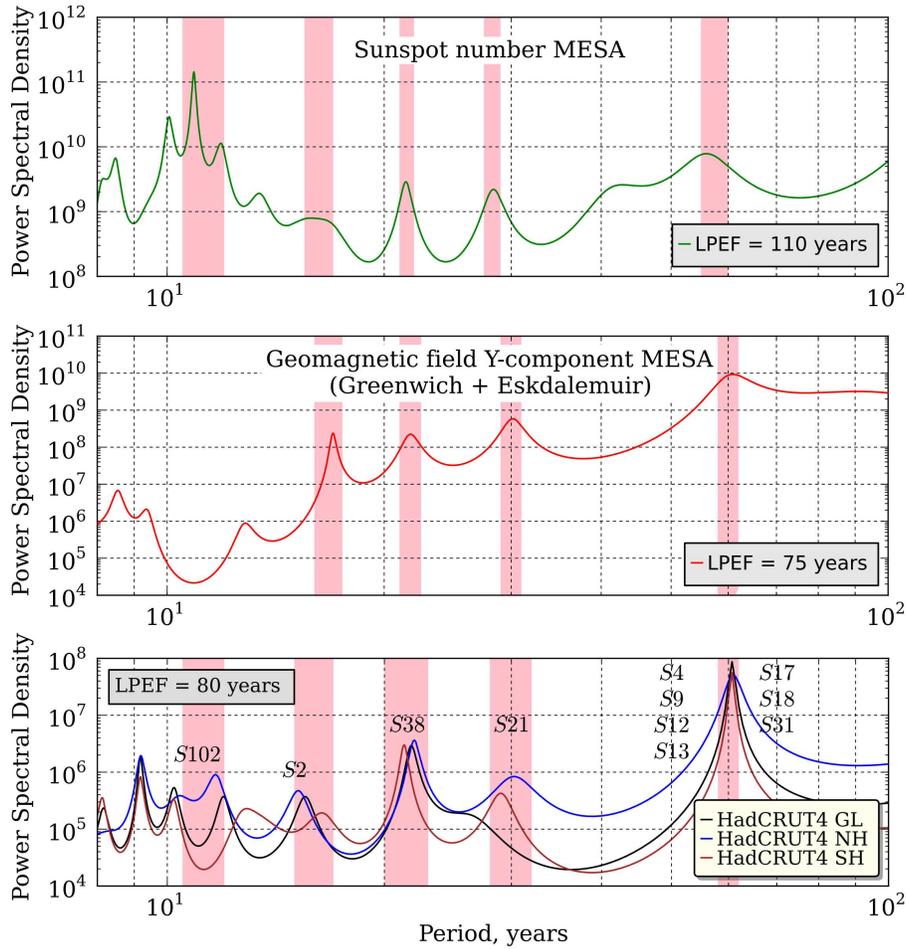}\end{center}
\caption{Power spectral densities for the sunspot number, the geomagnetic field Y-component and HadCRUT4 GST.}
\end{figure}

\

One can easily see that these harmonics may be associated with the orbits of the best known short-period S-stars (S102, S2, S38, S21,
S4-S9-S12-S13-S17-S18-S31) at the Galactic Center \cite{Gillessen,Schodel,Genzel} and with the solar cycles (about 11-12, 15-16, 20-22, 29-30, 60-61 years).
For example, the star S102 has the orbital period around $11.5$ years corresponding to the 11-year solar cycle (see the first pink stripe on the left).

\

In Fig. 2 we depict the power spectra of the global temperature \cite{Bintanja}. Again, pink stripes correspond to the major harmonics, which, in their turn,
may be associated with the orbits of the best known non-short-period S-stars (S19, S24, S66-S87, S97, S83, S?) at the Galactic Center
\cite{Gillessen,Schodel,Genzel} and with the solar cycles (about 250, 330, 500, 1050, 1700, 3600 years). For example, the star S19 has the orbital period
$(260\pm31)$ years presumably corresponding to the 250-year solar cycle (see the first pink stripe on the left). Green stripes represent the major temperature
oscillations, which are presumably associated with the variations of the Earth orbital parameters: eccentricity ($\sim$93~kyr), obliquity ($\sim$41~kyr) and
axis precession ($\sim$23~kyr).

\

\begin{figure}[htbp]
\begin{center}\includegraphics[width=4.85in,height=3.60in]{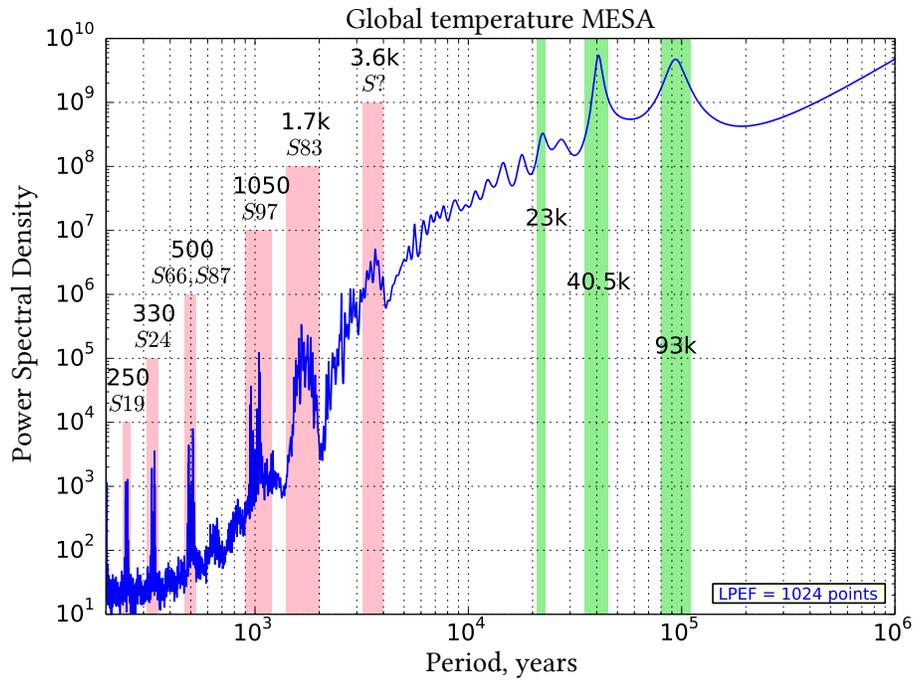}\end{center}
\caption{Power spectral density for the global temperature.}
\end{figure}

\

Thus, we see that the answer to the raised question is definite and positive: the compared periods coincide with each other.


\section*{Conclusion}

In this brief Letter on basis of the experimental data on the oscillations of the sunspot number, the geomagnetic field Y-component and the global temperature
we have explicitly demonstrated that their periods coincide with revolution periods of S-stars orbiting a supermassive black hole at the Galactic Center of the
Milky Way. It is absolutely obvious that such a fine coincidence cannot be random. Then the next quite natural question arises: how do the solar and
terrestrial observables "know" about motion of S-stars? And a hypothesis inevitably comes to mind: the "carrier" is none other than dark matter. More
specifically, S-stars can modulate dark matter flows in our galaxy and, consequently, cause variations of dark matter space and velocity distributions, in
particular, at the Sun and Earth positions. Further, these variations may cause the corresponding variations of the Solar System observables by means of some
mechanism, e.g., the interaction of dark matter particles with the cores of the Sun and the Earth. Such a probable mechanism as well as the above-mentioned
modulation itself are beyond the scope of our Letter and require a separate profound investigation, which is a subject of the forthcoming paper. Here our aim
is to stress that the available experimental data indicate the frequency transfer from the center of our galaxy to the Solar System. This fact can serve as an
indirect evidence of the proposed hypothesis that dark matter plays the role of the variations carrier.


\section*{Acknowledgements}

The work of M. Eingorn was supported by NSF CREST award HRD-1345219 and NASA grant NNX09AV07A.



\begin{thebibliography}{}
%
\bibitem{Fernandez}
E. Fernandez-Martinez and R. Mahbubani, {\em The Gran Sasso muon puzzle}, JCAP {\bf 07} (2012) 029; arXiv:astro-ph/1204.5180.
%
%
\bibitem{Gillessen}
S. Gillessen, F. Eisenhauer, S. Trippe, T. Alexander, R. Genzel, F. Martins, and T. Ott, {\em Monitoring stellar orbits around the massive black hole in the
galactic center}, The Astrophysical Journal {\bf 692} (2009) 1075; arXiv:astro-ph/0810.4674.
%
%
\bibitem{Schodel}
R. Sch\"odel, A. Feldmeier, N. Neumayer, L. Meyer, and S. Yelda, {\em The nuclear cluster of the Milky Way: our primary testbed for the interaction of a dense
star cluster with a massive black hole}; arXiv:astro-ph/1411.4504.
%
%
\bibitem{Genzel}
R. Genzel, F. Eisenhauer, and S. Gillessen, {\em The galactic center massive black hole and nuclear star cluster}, Reviews of Modern Physics {\bf 82} (2010)
3121; arXiv:astro-ph/1006.0064.
%
%
\bibitem{Bintanja}
R. Bintanja and R.S.W. van de Wal, {\em North American ice-sheet dynamics and the onset of 100,000-year glacial cycles}, Nature {\bf 454} (2008) 869.
%
\end{thebibliography}
\end{document}